# Near-infrared emission from spatially indirect excitons in type II ZnTe/CdSe/(Zn,Mg)Te core/double-shell nanowires

Piotr Wojnar[1*], Jakub Płachta[1], Anna Reszka[1], Jonas Lähnemann[3], Anna Kaleta[1], Sławomir Kret[1], Piotr Baranowski[1], Maciej Wójcik[1], Bogdan J. Kowalski[1], Lech T. Baczewski[1], Grzegorz Karczewski[1], Tomasz Wojtowicz[1,2]

[1] Institute of Physics, Polish Academy of Sciences, Al Lotników 32/46, PL-02-668 Warsaw, Poland
[2] International Research Centre MagTop, Institute of Physics, Polish Academy of Sciences, Al Lotników 32/46, PL-02-668 Warsaw, Poland
[3] Paul-Drude-Institut für Festkörperelektronik, Leibniz-Institut im Forschungsverbund Berlin e.V., Hausvogteiplatz 5-7, 10117 Berlin, Germany

E-mail: wojnar@ifpan.edu.pl

**Abstract**

ZnTe/CdSe/(Zn,Mg)Te core/double-shell nanowires are grown by molecular beam epitaxy by employing the vapor-liquid-solid growth mechanism assisted with gold catalysts. A photoluminescence study of these structures reveals the presence of an optical emission in the near infrared. We assign this emission to the spatially indirect exciton recombination at the ZnTe/CdSe type II interface. This conclusion is confirmed by the observation of a significant blue-shift of the emission energy with an increasing excitation fluence induced by the electron-hole separation at the interface. Cathodoluminescence measurements reveal that the optical emission in the near infrared originates from nanowires and not from two dimensional residual deposits between them. Moreover, it is demonstrated that the emission energy in the near infrared depends on the average CdSe shell thickness and the average Mg concentration within the (Zn,Mg)Te shell. The main mechanism responsible for these changes is associated with the strain induced by the (Zn,Mg)Te shell in the entire core/shell nanowire heterostructure.

## 1. Introduction

The staggered band alignment characteristic for a type II semiconductor heterostructure leads to a spatial separation of conduction band electrons and valence band holes at the interface. This effect can be employed, for instance, in photovoltaics where the electron hole separation process is crucial for the device performance and is usually obtained by the presence of a p-n junction. Moreover, the spatially indirect optical transition usually appears at energies smaller than the energy gaps of both semiconductors constituting the junction. This fact expands significantly the family of semiconductors that are useful for photovoltaic applications. In particular, it enables the use of wide band semiconductors for this purpose which has been demonstrated for ZnO/ZnSe [1–4], ZnO/ZnS [5–7], ZnO/ZnTe[8,9], ZnO/CdS [10], ZnSe/CdS [11], CdSe/ZnTe [12], InAs/In(As,Sb) [13] coaxial nanowire heterostructures. Another important motivation for studying the carrier separation at the type II nanowire (NW) heterojunctions is the opportunity to investigate coherently rotating states, which should be manifested by the presence of Aharonov-Bohm oscillations in an external magnetic field [14–16]. The latter effect might be employed in the framework of several protocols proposed for quantum information storage [17–19].

The CdSe/ZnTe semiconductor system is particularly well suited for the formation of heterostructures due to a relatively small lattice mismatch of 0.3%. CdSe/ZnTe core/shell nanowires with wurtzite crystal structure have already been used as building blocks of photovoltaic devices [12], as well as in photodetectors with the performance enhanced by the phototronic effect [20]. While the bandgaps of ZnTe and CdSe are equal to 2.39 eV and 1.75 eV, respectively, the emission at the type II interface is expected to appear at about 1.01 eV, i.e. in the near-infrared spectral region. Up to date, the near-infrared emission from CdSe/ZnTe core/shell NWs has not been reported. The influence of the type II character of the ZnTe/CdSe core/shell interface on the optical emission has been shown to be manifested by a significant decrease of the core-related excitonic emission [12]. In other papers, the emission energy variation as a function of the chemical composition is reported for highly luminescent (Zn,Cd)Te/CdSe colloidal core/shell NWs [21] and Cd(Se,Te)/ZnTe self-assembled quantum dots [22]. However, the luminescence always appears in the visible spectral range. Only a weak near-infrared optical emission has been observed in the case of planar CdSe/ZnTe heterojunctions [23].

In this work, the fabrication of ZnTe/CdSe/(Zn,Mg)Te core/double shell nanowire heterostructures is reported. The vapor-liquid-solid growth mechanism assisted by gold catalysts in a system for molecular beam epitaxy (MBE) is employed. A schematic picture of the cross-section of the studied NWs is shown in Figure 1a. First of all, it is demonstrated that an optical emission appears in the near infrared spectral range with the maximum varying from 1.03 eV to 1.17 eV depending on the parameters of the particular heterostructure and on the excitation fluence. A combined study performed by means of photoluminescence (PL) and cathodoluminescence (CL) shows unambiguously that this emission originates, indeed, from the type II CdSe/ZnTe core/shell interface within the NWs, as marked with arrows in the schematic picture in Figure 1b. That is why the studied NW heterostructures might be potentially used not only for solar energy conversion applications, but also as a building block of near-infrared light emitting devices. Such devices could be applied in the field of information and communications technology [26–29] taking advantage of the fact that the light from this spectral range is nearly invisible, as well as in the field of biomedical imaging purposes [30,31] thanks to the semitransparency of human tissue to the light from this spectral range.

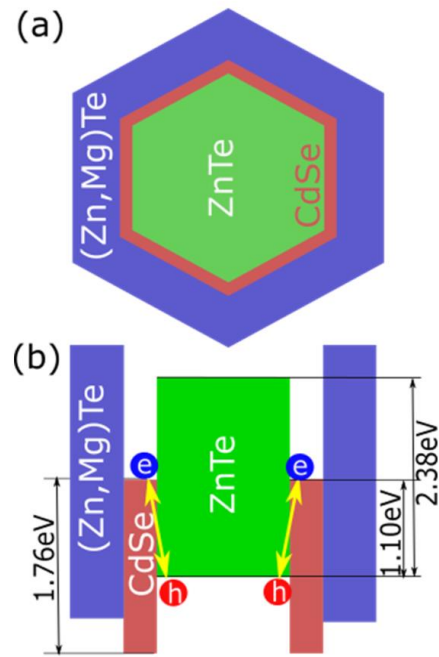

**Figure 1.** (a) Schematic cross-section of a ZnTe/CdSe/(Zn,Mg)Te core/double shell NW (b) Band diagram of the heterostructure. Electrons tend to localize within the CdSe shell and holes within ZnTe core. The optical transition at the type II ZnTe/CdSe interface is marked with arrows. Band gap values are based on ref [24] and the CdSe/ZnTe band offset is taken from [25].

## 2. Growth and structural properties

ZnTe/CdSe/(Zn,Mg)Te core/double shell nanowire heterostructures, which are the subject of this report, are grown in the MBE system consisting of two growth chambers coupled by an ultra-high vacuum connection, dedicated to the growth of either tellurium based (EPI 620 system) or selenium based semiconductors (Prevac). The first chamber is equipped with Te, Zn and Mg effusion cells and atomic fluxes given by the beam equivalent pressure (BEP) equal to $1.1 \times 10^{-6}$, $4.5 \times 10^{-7}$ and $3.0 \times 10^{-8}$ Torr, respectively, whereas in the second chamber Cd, Se fluxes are set to have BEP values equal to $7.5 \times 10^{-7}$ and $1.2 \times 10^{-6}$ Torr. At the first stage, ZnTe NWs are grown on (111) oriented silicon substrates by employing the vapor-liquid-solid growth mechanism assisted by gold catalysts following the procedure described in ref [32]. ZnTe NW growth takes place at 410 °C for 35 min. A typical length of ZnTe NWs is about 1.0 µm and their diameters range from 20 to 40 nm. At the next step, the sample is transferred to the second MBE chamber for the CdSe shell deposition. After stabilizing the temperature at 300 °C in a Cd flux, the shell is deposited for 3 or 6 min, depending on the particular sample, whereas the approximate growth rate amounts to 0.7 nm/min as determined later by electron microscopy. An important parameter for CdSe deposition is the growth temperature, which should be in the range of 290 °C - 320 °C. At temperatures above 330 °C, the shell is probably not deposited, as inferred from the absence of any optical emission from the CdSe/ZnTe interface in any sample grown under these conditions. Finally, the sample is transferred back to the first growth chamber for the deposition of the (Zn,Mg)Te passivation shell. After stabilizing the substrate temperature at 350 °C under a Te flux, the growth takes place during 10 minutes with an average growth rate of about 2 nm/min [33] leading to an average shell thickness of about 20 nm. Mg concentration is determined subsequently by energy dispersive X-ray spectroscopy (EDS) measurements and amounts to 0.25 in most of the samples, unless stated differently.

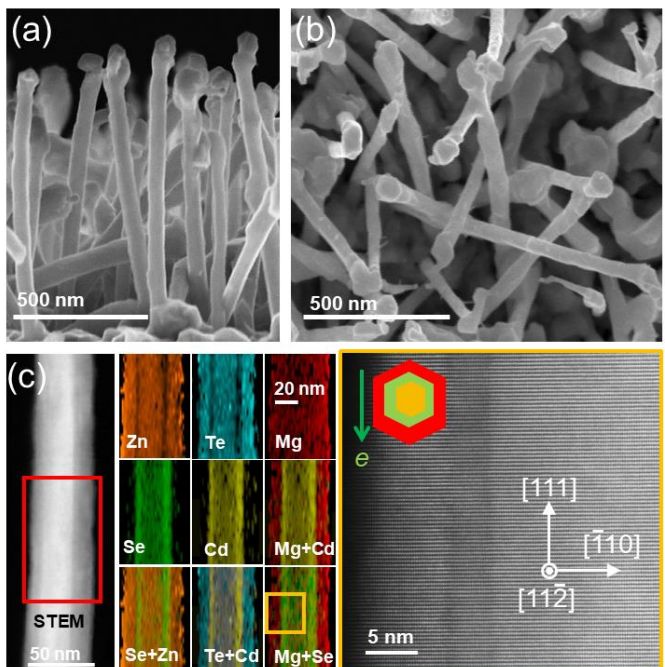

**Figure 2** Morphology and elemental composition of ZnTe/CdSe/(Zn,Mg)Te core/double-shell NWs. SEM images of 'as grown' NWs, acceleration voltage 5 kV, probe current 40 pA (a) side view (b) top view; (c) EDS maps performed in a TEM revealing the distribution of Zn, Mg, Te, Cd, Se atoms within the NW (d) Crystalline structure of the core and both shells measured by STEM with high spatial resolution along the [11-2] zone axis.

The NW morphology has been studied by means of scanning electron microscopy (SEM, Zeiss Auriga). The electron beam acceleration voltage is 5 kV and the probe current - 40 pA. The average NW length is found to be 0.9 ± 0.2 μm whereas the overall NW diameter is 60 ± 8 nm as determined from side-view SEM images shown in Figure 2a. Moreover, the NWs are oriented at random directions with respect to the substrate, which is clearly visible in Figure 2b. The absence of an epitaxial relation between the NWs and the (111) oriented Si substrate is, most likely, due to a residual silicon oxide layer which is not completely desorbed from the substrate [32]. The spatial distribution of Zn, Mg, Te, Cd and Se elements within a NW is examined with the use of an EDS detector inside a Titan Cube 80-300 transmission electron microscope (TEM) operating at 300 kV in scanning transmission electron microscopy (STEM) mode (Figure 2c). The EDS detector active area amounts to 30 mm$^2$ and its solid angle equals to 0.13 srad. The beam current is set to 200 pA which corresponds to the size of the electron probe given by the full width at half maximum (FWHM) of the electron density distribution smaller than 0.2 nm. For the measurements performed with high spatial resolution the beam current is reduced to 60 pA and the size of the electron probe is smaller than 0.13 nm. For EDS element mapping the NW which has the shape of a hexagonal pencil is rotated so that the two side walls as well as shell interfaces are parallel to the incident electron beam, as shown in the inset of Figure 2d. In such orientation electron beam is parallel to [11-2] zone axis. It is found that the signals from the Zn, Te and Mg are detectable within the entire NW, however with the varying intensity. On the other hand, strong Cd and Se signals come only from a thin shell, which surrounds the ZnTe core. Therefore, their distribution reflects the shape of the NW core with the characteristic diameter of about 30 nm, as seen in Figure 2c. In order to obtain insights into the crystalline structure, TEM measurements with high spatial resolution (HRTEM) have been performed (Figure 2d). It is found that the nanowire heterostructure crystallizes entirely in an almost perfect defect free zinc blende structure with <111> direction being always the NW axis. Most importantly, the epitaxial relation is conserved between the core and both shells. This fact is confirmed by the STEM images of the NW cross-sections presented in Figure 3. Any stacking faults or twin boundaries are not found in the images taken in the <110> zone axis (not shown). The excellent crystalline properties of the nanowires are an effect of the optimization of the ZnTe nanowires growth parameters, such as the Zn/Te flux ratio and the growth temperature. The epitaxial relation between the core and both shells is maintained due to the relatively small lattice mismatch between the core and shell semiconductors and due to the fact that the shell thickness do not exceed critical values for the generation of misfit dislocations.

In Figure 3a, one can clearly distinguish the ZnTe core characterized by a hexagonal shape as well as both the CdSe and (Zn,Mg)Te shells. In this particular nanowire the shell thickness is found to depend on the angular orientation. For the CdSe shell, it varies from 2 to 6 nm, whereas for the (Zn,Mg)Te shell from 5 to 16 nm. Most likely, this variation of the thickness is caused by a shadowing effect taking place during the MBE growth of the shell and depends on the spatial orientation of the particular nanowire with respect to the atomic fluxes. The close-up at the CdSe shell presented in Figure 3b confirms the epitaxial relation between the core and both shells.

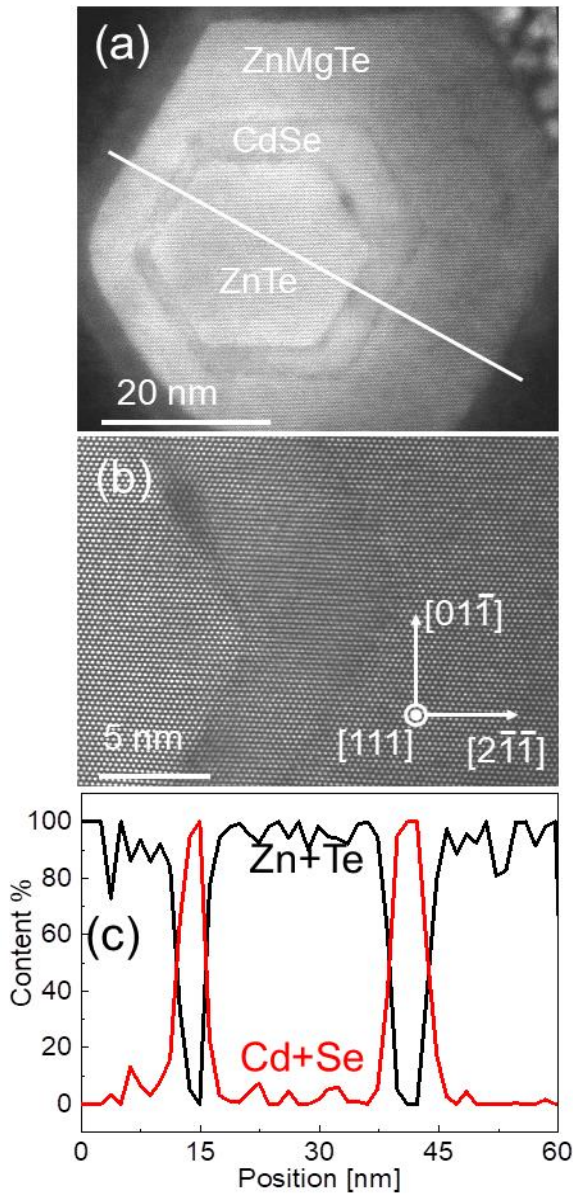

**Figure 3** (a) STEM images of the ZnTe/CdSe/(Zn,Mg)Te core/double-shell nanowire cross-section revealing the shape and the thicknesses of both shells (b) close-up indicating an epitaxial relation between the core and both shells. (c) EDS plot of the sum of Cd+Se and Zn+Te signals along the white line shown in (a). The scale bar in (b) is 5 nm.

Moreover, a thin interfacial layer with the thickness of about 1 nm, presumably composed of a mixed (Zn,Cd)(Se,Te) phase, is present at the core/shell and shell/shell interfaces. On the basis of the STEM images, Figure 2d and 3b, it is not possible to unequivocally determine the degree of mixing of these elements, which may result either from the interdiffusion effects or from the roughness of the interfaces. However, it is unlikely that this interfacial layer is composed of pure ZnSe, as reported previously for planar CdSe/ZnTe interfaces [34].

EDS element profiles presented in Figure 3c confirm the chemical composition of the core and both shells, which are ZnTe, CdSe and (Zn,Mg)Te with Mg content of 0.25, respectively. In particular, it is found that at the spatial position where CdSe is present, the ZnTe signal decreases significantly. However, from these measurements no conclusions on the composition of a thin interfacial layer can be drawn due to the insufficient spatial resolution of the EDS technique.

## 3. Optical emission

For photoluminescence measurements the samples are mounted within a closed cycle hydrogen cryostat at 7 K. The excitation is performed with a 532 nm laser whereas the excitation spot size is of the order of 0.1 mm. That is why tens of thousands nanowires are measured simultaneously. The detection system consists of a 303 cm monochromator (SR303i by Andor) with a 300 groves/mm grating blazed for 1500 nm equipped with an iDus 1.7 μm InGaAs detector array.

The optical emission spectrum from 'as grown' samples consists typically of two emission bands, the low energy band with maximum at 1.0-1.2 eV and the high energy band at 1.70-1.80 eV (see Figure 4a). The high energy band corresponds well to the energy gap of CdSe. However, as demonstrated subsequently by cathodoluminescence studies, it does not originate from the NWs themselves but from residual cluster-like deposits present on the substrate between the NWs. On the other hand, the emission energy of the low energy band fits to the expected value of the spatially indirect transition at the type II ZnTe/CdSe interface [23]. In Figure 4a, the low temperature PL spectra collected from three different samples containing ZnTe/CdSe/(Zn,Mg)Te core/double-shell NWs are presented. The maximum energy varies from 1.09 eV to 1.17 eV, i.e. by 80 meV, depending on the shell parameters, such as the CdSe shell thickness and Mg content in the (Zn,Mg)Te shell. The small emission peak at 1.07 eV comes from silicon substrate and is visible in samples with a weak emission from the nanowires.

Let us first consider the influence of the CdSe shell thickness on the emission energy by comparing the spectra (1) and (2) from Figure 4a. With an increasing CdSe shell thickness, one observes a blue-shift of the emission energy. This is a quite unexpected effect, since the quantum size effect should have an opposite impact on the emission energy. Therefore, our explanation relies on the strain present within the heterostructure.

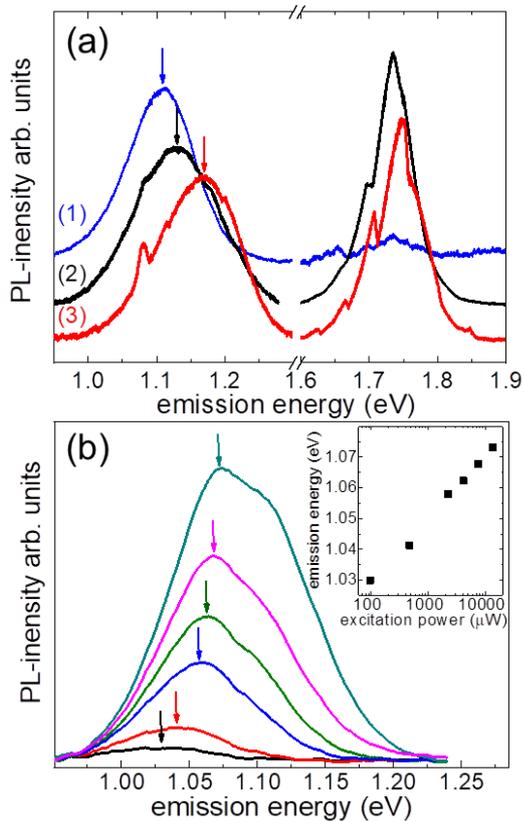

**Figure 4** Photoluminescence of ZnTe/CdSe/(Zn,Mg)Te core/double-shell NWs. (a) Spectra from three samples with different CdSe average shell thickness, d, and Mg content in (Zn,Mg)Te, $x_{Mg}$ (1) d = 2 nm, $x_{Mg}$ = 0.25 (2) d = 4 nm, $x_{Mg}$ = 0.25 (3) d = 4 nm, $x_{Mg}$ = 0.17 (b) Power dependence of the PL for excitation powers in the range from 100 μW to 13.3 mW. Inset: the increase of emission energy as a function of excitation power. Temperature of the measurement is 7 K, excitation laser line – 532 nm (2.33 eV).

As already well established for lattice mismatched core/shell nanowire heterostructures, the strain conditions within the core are determined by the core/shell lattice mismatch and the ratio of the core to the entire NW diameter [35–38]. The (Zn,Mg)Te shell induces a significant tensile stress in the ZnTe core. The larger the Mg concentration in the shell, the larger is the strain present within the structure (for a Mg concentration of 0.25, the core/shell lattice mismatch amounts to 13% and for a Mg concentration of 0.17 to 9%.). On the other hand, the CdSe/ZnTe lattice mismatch has an opposite sign and amounts to only - 0.3%. Therefore, it may be assumed, to a good approximation, that the strain in the system originates mostly from the lattice mismatched (Zn,Mg)Te shell. Moreover, the presence of a CdSe shell should reduce the tensile stress acting on the core. As already well established [33], the tensile stress acting on the ZnTe NW core induces a decrease of the band gap resulting in a red-shift of the excitonic emission. Therefore, the reduction of this stress by the increase of the CdSe shell thickness should result in a blue-shift of the excitonic emission, as observed in Figure 4a. The impact of the strain on the emission energy is also manifested in the dependence of the emission energy on the Mg concentration within the (Zn,Mg)Te shell, i.e., when comparing spectra (1) and (3) in Figure 4a. It is found that the near-infrared emission from the NWs with lower Mg content (3) appears at higher energies compared to the emission from structures with higher Mg content (1). This fact is consistent with our interpretation in terms of a strain-induced bandgap variation since the emission from the less strained NW heterostructure appears at higher energies.

At this stage, it is important to show that the emission in the near infrared originates from the CdSe/ZnTe interface and not from any deep level defects in one of the semiconductors constituting the junction. For this purpose, excitation fluence dependent measurements of the optical emission spectra are performed. Importantly, a pronounced blue-shift of the emission is observed as a function of the excitation fluence, as shown in Figure 4b. This effect is characteristic for the spatially indirect emission at a type II interface and has been observed in several semiconductor systems including type II quantum wells [23] or monolayer $MoSe_2$–$WSe_2$ heterostructures [39]. The reason for the blue-shift is the creation of electric dipoles caused by the spatial separation of electrons and holes at the interface. The increase of the excitation fluence leads to an accumulation of these dipoles at the type II interface. The increase of the repulsive Coulomb dipole-dipole interaction leads to the overall blue-shift of the emission energy.

After demonstration that the emission in the near-infrared range originates from the type II ZnTe/CdSe interface, it has to be clarified that it comes only from the NWs and not from any other structures, which are produced simultaneously with the growth of NWs, such as e.g., the two-dimensional residual cluster-like deposits in between the NWs or 'crooked' NWs. For this purpose, CL measurements are performed using a Zeiss Ultra55 SEM equipped with a Gatan MonoCL4 system for CL signal collection, a nitrogen-cooled infrared-photomultiplier for CL-signal detection and a helium-cooled sample stage. The temperature of the measurement is stabilized at 7 K, the acceleration voltage set to 5 kV, and the probe current to 20 nA.

The CL spectrum of an 'as grown' sample consists of two emission bands with the maximum at 1.04 eV and at about 1.73 eV (Figure 5a), which agrees with the PL spectrum presented in Figure 4. In order to confirm that these emission peaks originate from the NWs, monochromatic mapping of the CL-signal has been performed at several wavelengths selected by means of an optical spectrometer, in particular at 1250 nm (0.99 eV) and 716 nm (1.73 eV). The SEM image of the studied area reveals the presence of several tens of NWs, as shown in Figure 5b. It is found that only for the CL mapping at 1250 nm depicted in Figure 5b', the shapes and spatial positions of the NWs are quite well reproduced, despite of the fact that the signal is weak. This finding leads us to the conclusion that the emission at this wavelength comes, indeed, from the NWs.

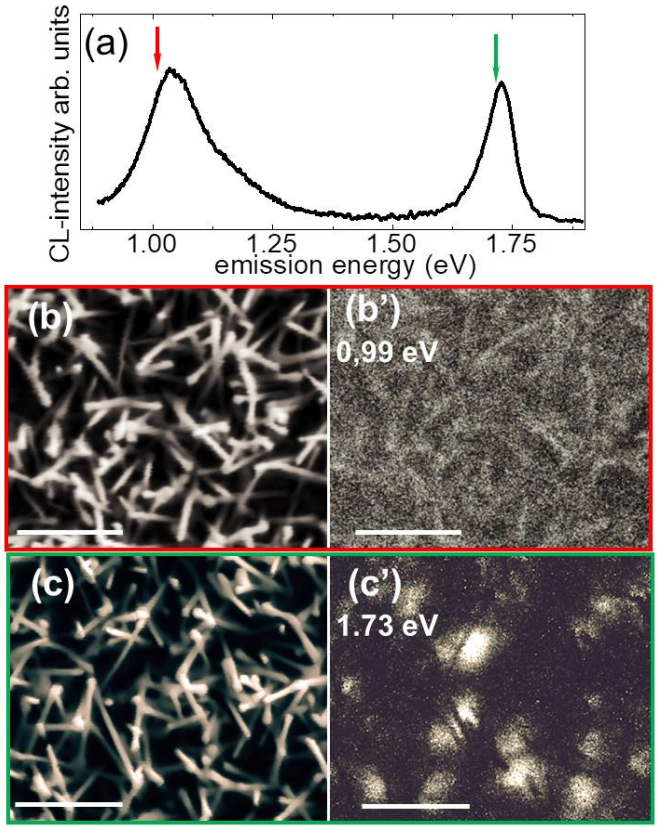

**Figure 5.** CL from ZnTe/CdSe/(Zn,Mg)Te core/double-shell NWs with d = 4 nm and $x_{Mg}$ = 0.25 (a) CL spectrum (b) SEM image and (b') monochromatic CL mapping at 1250 nm (0.99 eV) of the same area. (c) SEM image and (c') CL-mapping at 716 nm (1.73 eV). The temperature of the measurement is 7 K, the acceleration voltage 5 kV, and the probe current 20 nA. The scale bars correspond to 1 μm.

On the contrary, no analogous conclusion can be drawn from the CL maps performed at 716 nm, Figures 5c and 5c'. The size of the visible bright emitting areas is significantly larger than the NW diameter. The shapes and locations of the emitting objects definitely do not correspond to NWs but rather to the NWs free, exposed regions of the substrate. Therefore, we conclude that this emission band is attributed to residual CdSe deposits present between the NWs. The latter finding underlines the importance of the use of highly spatially resolved CL measurements for the identification of the origin of the optical emission in such nanostructures.

## 4. Conclusions

High quality ZnTe/CdSe/(Zn,Mg)Te core/double-shell NWs are grown by MBE on (111) oriented silicon substrates by applying the gold-assisted vapor-liquid-solid growth mechanism. The NWs are characterized by an almost perfect defect free zincblende crystal structure with the nanowire axis always being the <111> direction. Optical emission spectra of these structures consist of two emission bands at about 1.1 eV and 1.75 eV. The emission energy of the low energy band exhibits a significant blue-shift with increasing excitation fluence, which leads us to associate it to the emission from spatially indirect excitons at the type II ZnTe/CdSe interface. CL measurements confirm that the emission in the near-infrared region originates from the NWs, whereas the high energy band comes from residual CdSe deposits formed between the NWs. The energy of the low energy band can be tuned as much as 80 meV by varying the thickness of the CdSe shell and Mg content in the outer shell. The main mechanism of these changes is identified to be related to the strain induced by (Zn,Mg)Te shell in the entire heterostructure and not to the quantum size effect.

Based on these results, we conclude that the optical emission in the near infrared is related to the recombination of spatially indirect excitons at the type II ZnTe/CdSe interface. The presence of optical transitions in the near-infrared spectral region, the spontaneous carrier separation at the type II interface and the relatively small lattice mismatch between these two semiconductors combined with the high surface-to-volume ratio typical for NWs makes ZnTe/CdSe core/shell NWs a prospective nanostructure for the application in photovoltaic devices.

### Acknowledgements

This work has been partially supported by the National Centre of Science (Poland) through grant 2017/26/E/ST3/00253, and by the Foundation for Polish Science through the IRA Programme co-financed by the EU within SG OP (grant No. MAB/2017/1).

### References

[1]  Oksenberg E, Marti-Sanchez S, Popovitz-Biro R, Arbiol J and Joselevich E 2017 *ACS Nano* **11** 6155–66

[2]  Jabri S, Amiri G, Hassani S, Lusson A, Sallet V, Meftah A, Galtier P and Oueslati M 2017 *Appl. Phys. Lett.* **110**

[3]  Xiao C H, Wang Y D, Yang T Y, Luo Y and Zhang M Z 2016 *Appl. Phys. Lett.* **109**

[4]  Zheng J J, Wu Z M, Yang W H, Li S P and Kang J Y 2010 *J. Mater. Res.* **25** 1272–7

[5] Brayek A, Ghoul M, Souissi A, Ben Assaker I, Lecoq H, Nowak S, Chaguetmi S, Ammar S, Oueslati M and Chtourou R 2014 *Mater. Lett.* **129** 142–5

[6] Liu S, Wang X T, Zhao W X, Wang K, Sang H X and He Z 2013 *J. Alloys Compd.* **568** 84–91

[7] Wen S W, Huang B H, Kang J J, Lin T Y, Hsu H F, Ni H C, Hou C H, Lai H T and Gong J R 2016 *Ecs J. Solid State Sci. Technol.* **5** 663–6

[8] Schrier J, Demchenko D O and Wang L W 2007 *Nano Lett.* **7** 2377–82

[9] Chao H Y, Cheng J H, Lu J Y, Chang Y H, Cheng C L and Chen Y F 2010 *Superlattices Microstruct.* **47** 160–4

[10] Yang X, Li H, Zhang W, Sun M X, Li L Q, Xu N, Wu J D and Sun J 2016 *Appl. Phys. Lett.* **109**

[11] Zhang X W, Mao J, Shao Z B, Diao S L, Hu D, Tang Z J, Wu H H and Jie J S 2017 *J. Mater. Chem. C* **5** 2107–13

[12] Wang K, Rai S C, Marmon J, Chen J J, Yao K, Wozny S, Cao B B, Yan Y F, Zhang Y and Zhou W L 2014 *Nanoscale* **6** 3679–85

[13] Ercolani D, Gemmi M, Nasi L, Rossi F, Pea M, Li A, Salviati G, Beltram F and Sorba L 2012 *Nanotechnology* **23**

[14] Corfdir P, Marquardt O, Lewis R B, Sinito C, Ramsteiner M, Trampert A, Jahn U, Geelhaar L, Brandt O and Fomin V M 2019 *Adv. Mater.* **31**

[15] Sellers I R, Whiteside V R, Kuskovsky I L, Govorov A O and McCombe B D 2008 *Phys. Rev. Lett.* **100**

[16] Kuskovsky I L, Mourokh L G, Roy B, Ji H, Dhomkar S, Ludwig J, Smirnov D and Tamargo M C 2017 *Phys. Rev. B* **95**

[17] Fischer A M, Campo V L, Portnoi M E and Romer R A 2009 *Phys. Rev. Lett.* **102**

[18] Climente J I and Planelles J 2014 *Appl. Phys. Lett.* **104** 193101

[19] Simonin J, Proetto C R, Pacheco M and Barticevic Z 2014 *Phys. Rev. B* **89** 75304

[20] Rai S C, Wang K, Ding Y, Marmon J K, Bhatt M, Zhang Y, Zhou W L and Wang Z L 2015 *ACS Nano* **9** 6419–27

[21] Groeneveld E, Van Berkum S, Van Schooneveld M M, Gloter A, Meeldijk J D, Van Den Heuvel D J, Gerritsen H C and De Mello Donega C 2012 *Nano Lett.* **12** 749–57

[22] Baranowski P, Szymura M, Karczewski G, Aleszkiewicz M, Rodek A, Kazimierczuk T, Kossacki P, Wojtowicz T, Kossut J and Wojnar P 2020 *Appl. Phys. Lett.* **117**

[23] Mourad D, Richters J P, Gerard L, Andre R, Bleuse J and Mariette H 2012 *Phys. Rev. B* **86**

[24] Adachi S 2005 *Properties of Group-IV, III-V and II-VI Semiconductors* (Chichester, UK: John Wiley & Sons, Ltd)

[25] Wei S H and Zunger A 1998 *Appl. Phys. Lett.* **72** 2011–3

[26] Ning J, Xiong Y, Huang F, Duan Z, Kershaw S V. and Rogach A L 2020 *Chem. Mater.* **32** 7842–9

[27] Bai X, Purcell-Milton F and Gun'ko Y K 2020 *Nanoscale* **12** 15295–303

[28] Van Der Stam W, Bladt E, Rabouw F T, Bals S and De Mello Donega C 2015 *ACS Nano* **9** 11430–8

[29] Sayevich V, Robinson Z L, Kim Y, Kozlov O V., Jung H, Nakotte T, Park Y S and Klimov V I 2021 *Nat. Nanotechnol.* **16** 673–9

[30] Gu Y, Guo Z, Yuan W, Kong M, Liu Y, Liu Y, Gao Y, Feng W, Wang F, Zhou J, Jin D and Li F 2019 *Nat. Photonics* **13** 525–31

[31] Liang L, Feng Z, Zhang Q, Cong T Do, Wang Y, Qin X, Yi Z, Ang M J Y, Zhou L, Feng H, Xing B, Gu M, Li X and Liu X 2021 *Nat. Nanotechnol.* 1–6

[32] Wojnar P, Szymura M, Zaleszczyk W, Klopotowski L, Janik E, Wiater M, Baczewski L T, Kret S, Karczewski G, Kossut J and Wojtowicz T 2013 *Nanotechnology* **24** 365201

[33] Wojnar P, Zielinski M, Janik E, Zaleszczyk W, Wojciechowski T, Wojnar R, Szymura M, Klopotowski L, Baczewski L T, Pietruchik A, Wiater M, Kret S, Karczewski G, Wojtowicz T and Kossut J 2014 *Appl. Phys. Lett.* **104**

[34] Bonef B, Gérard L, Rouvière J L, Grenier A, Jouneau P H, Bellet-Amalric E, Mariette H, André R and Bougerol C 2015 *Appl. Phys. Lett.* **106** 051904

[35] Raychaudhuri S and Yu E T 2006 *J. Appl. Phys.* **99**

[36] Ferrand D and Cibert J 2014 *EPJ Appl. Phys.* **67**

[37] Artioli A, Rueda-Fonseca P, Stepanov P, Bellet-Amalric E, Den Hertog M, Bougerol C, Genuist Y, Donatini F, André R, Nogues G, Kheng K, Tatarenko S, Ferrand D and Cibert J 2013 *Appl. Phys. Lett.* **103**

[38] Trammell T E, Zhang X, Li Y, Chen L-Q and Dickey E C 2008 *J. Cryst. Growth* **310** 3084–92

[39] Rivera P, Schaibley J R, Jones A M, Ross J S, Wu S, Aivazian G, Klement P, Seyler K, Clark G, Ghimire N J, Yan J, Mandrus D G, Yao W and Xu X 2015 *Nat. Commun.* **6** 1–6